\documentclass[11pt,a4paper]{article}
\usepackage[utf8]{inputenc} 
\usepackage[T1]{fontenc}    
\usepackage[hyphens,spaces,obeyspaces]{url}
\usepackage[colorlinks=true,linkcolor=blue,citecolor=blue,urlcolor=blue]{hyperref}  
\usepackage{graphicx}
\usepackage{caption}
\usepackage{pdfpages}
\usepackage{float}
\usepackage{makecell}
\usepackage{amsmath,amssymb,amsthm,bm}
\usepackage{mathrsfs}
\usepackage{natbib}
\usepackage[normalem]{ulem}
\usepackage{comment}

\usepackage{anysize}
\usepackage{authblk}
\marginsize{25mm}{25mm}{25mm}{25mm}

\newcommand{\bX}{{\bm X}}

\newcommand{\bM}{{\bm M}}

\title{Relaxing door-to-door matching reduces passenger waiting times: a workflow for  the analysis of driver GPS traces in a stochastic carpooling service}

\author[1,2,3]{Panayotis Papoutsis\thanks{Corresponding author. E-mail: {\fontfamily{qcr}\selectfont papoutsispanayotis@gmail.com}}}
\author[3]{Safa Fennia}
\author[3]{Constant Bridon}
\author[3]{Tarn Duong}
 
\affil[1]{Department of Computing and Mathematics, Nantes Central Engineering School, F-44300, France}
\affil[2]{Jean Leray Mathematics Laboratory, University of Nantes, F-44300, France}
\affil[3]{Department of Data Science-GIS, Ecov, F-44200, France} 

\date{}

\begin{document}
\maketitle

\begin{abstract}
\noindent Carpooling has the potential to transform itself into a mass transportation mode by abandoning its adherence to deterministic passenger-driver matching for door-to-door journeys, and by adopting instead stochastic matching on a network of fixed meeting points. Stochastic matching is where a passenger sends out a carpooling request at a meeting point, and then waits for the arrival of a self-selected driver who is already travelling to the requested meeting point. Crucially there is no centrally dispatched driver. Moreover, the carpooling is assured only between the meeting points, so the onus is on the passengers to travel to/from them by their own means. Thus the success of a stochastic carpooling service relies on the convergence, with minimal perturbation to their existing travel patterns, to the meeting points which are highly frequented by both passengers and drivers. Due to the innovative nature of stochastic carpooling, existing off-the-shelf workflows are largely insufficient for this purpose. To fill the gap in the market, we introduce a novel workflow, comprising of a combination of data science and GIS (Geographic Information Systems), to analyse driver GPS traces. We implement it for an operational stochastic carpooling service in south-eastern France, and we demonstrate that relaxing door-to-door matching reduces passenger waiting times. Our workflow provides additional key operational indicators, namely the driver flow maps, the driver flow temporal profiles and the driver participation rates.
\medskip 

\noindent{Keywords: data science, stochastic matching, GIS, meeting point, network}
\end{abstract}

\section{Introduction}

Carpooling has seen an explosion of utilisation in recent years \citep{fuhurata2013}. There are many underlying reasons for this, with concerns ranging from greenhouse gas emissions and air pollution to road congestion to land use, as well as economic costs \citep{shaheen2016}. It also attracts intense interest since carpooling is a crucial element of almost all development plans for smart cities \citep{ghoseiri2012dynamic}. A broad definition of carpooling involves a driver sharing their journey with passengers. In this paper we employ a narrower definition. We additionally require that a non-professional driver would have undertaken their journey for their own reasons, regardless of whether the passengers would have been present or not. The driver may receive payment to offset the costs of the use of their vehicle, but the profit motive is non-existent or at least not their primary motivation \citep{zhu2020}. Hence we do not consider taxi-like services (such as Uber, Lyft and Kapten etc.) to be carpooling services as they employ professional drivers who create a journey in response to a passenger request and are then paid the market rate for the service rendered.  

Due to the altruistic nature of carpooling, service providers tend to be small, local, non-profit organisations. Though it does not preclude viable business models arising from carpooling with non-professional drivers: the market leader BlaBlaCar levies a commission fee for facilitating the matching of drivers and passengers \citep{shaheen2017}. This matching is managed by a centralised platform, which we call {\it deterministic matching} since a known driver is assigned in advance to collect the passenger. This deterministic matching is highly successful for infrequent, long distance, pre-reserved carpooling journeys, as witnessed by BlaBlaCar's status as a unicorn start-up company (a market capitalisation of at least 1000 million USD). Despite the success of deterministic passenger-driver matching in this market, attempts to export it other carpooling markets have not resulted in the same level of market penetration. This is most notable for frequent, short distance journeys (from 10 to 40 km roughly), which comprise the bulk of daily home-work commutes, and so carpooling remains a marginal practice in this market.

This paper focuses on short-distance, non-reserved carpooling, and it is what we refer to when employing `carpooling' without any qualifiers. The advent of mass carpooling depends crucially on incentivising drivers and passengers to converge onto highly frequented meeting points (hotspots) along their door-to-door journeys \citep{stiglic2015}.  This type of incentivisation is well-established for a bus network where passengers embark/disembark only at the fixed bus stops. Thus mass carpooling requires a paradigm shift from considering carpooling as an exclusively private means of transport to a closer alignment to public transport models \citep{cooper2007successfully}. 

Continuing with the public transport model, the meeting points are not defined informally between passengers and drivers, but are decided in consultation with local government authorities so that they respond to the mobility requirements in the local area, taking into account various factors such as aggregated traffic flow, socioeconomic characteristics, pedestrian accessibility, local government regulations, etc. For our purposes, we consider that the identification of the meeting hotspots has been carried out beforehand. These meeting points are then connected to each other to define carpooling lines, which have massification potential, like traditional bus lines \citep{stiglic2015,xin2018plosone}. 

Like a bus service, no pre-reservations are required, as a passenger makes an ad hoc  carpooling request at a meeting point, and this request for the desired destination is communicated to all passing drivers in real-time via an electronic sign on the side of a highly frequented main road.  Unlike for deterministic passenger-driver matching mentioned above, a specific driver is not assigned to the passenger by a centralised platform, but the decision to collect a passenger at the meeting point is made spontaneously by a self-selected driver. Since the actual driver who collects the passenger is not known deterministically in advance, but is only known to be drawn from the population of drivers, this is known as \textit{stochastic matching}. Due to the inherent variability of these driver arrivals, stochastic matching is only feasible when employed in conjunction with a network of highly frequented meeting points. 

The effects of the double innovations of fixed meeting points and stochastic matching are only sparsely covered by the recent comprehensive review of general carpooling and taxi-like services over the past two decades \citep{wang2019trb}. So there are few off-the-shelf workflows which are suitable for the analysis of the data arising from a stochastic carpooling service. We introduce a data science-GIS workflow which fills this gap in the market. Its main data source is the GPS traces, and its secondary sources are the meeting point locations, the origin-destination matrices, the route finder API and the base maps. Data wrangling/geoprocessing are then applied to these data sources, with the critical geoprocessing step being the topological simplification of the GPS traces onto the carpooling network. This topological simplification is essential to be able to mutualise GPS traces which share common arrival times at the meeting points. From these simplified GPS traces, we can produce the waiting times. The latter allow us to assert that stochastic matching at meeting points leads to reduced passenger waiting times in comparison to door-to-door matching. In addition to the waiting times, other outputs from this workflow are the driver flow maps, the driver flow profiles, and the driver participation rates. These additional outputs are obtained at low marginal cost but which are important elements for evidence-based decision making in a stochastic carpooling service. 

In Section~\ref{sec:door} we present the theoretical reasons why door-to-door matching is insufficient to ensure a regular carpooling service. In Section~\ref{sec:workflow} we detail our data science-GIS workflow for the analysis of GPS traces. In Section \ref{sec:case}, we apply this workflow to an operational stochastic carpooling service to produce the passenger waiting times and the other outputs. We end with some concluding remarks.

\section{Door-to-door matching is an obstacle to mass carpooling}\label{sec:door}

\begin{figure}[!ht]
\centering
\begin{tabular}{@{}c@{}c@{}}
\includegraphics[width=0.4\textwidth]{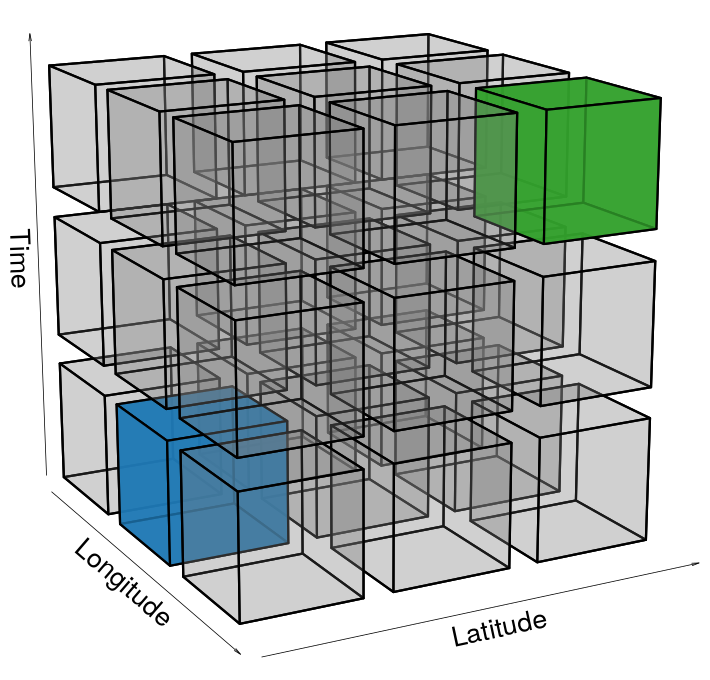} &
\includegraphics[width=0.4\textwidth]{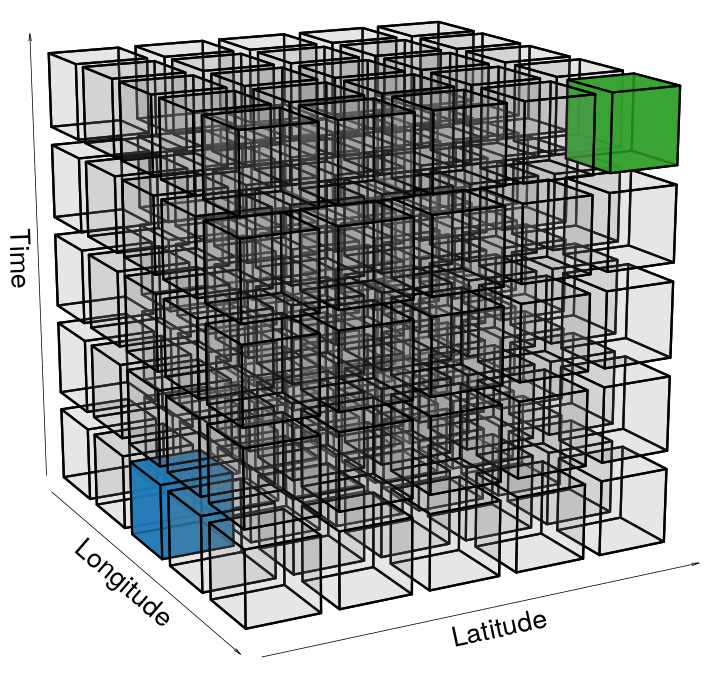}
\end{tabular}
\caption{Spatio-temporal door-to-door matching fragments the population of mutualisable trajectories. (Left) Relaxed matching conditions. (Right) Restricted matching conditions. Blue sub-cube represents the origin (residential neighbourhood), green the destination (workplace), and trajectories which share the same origin and destination sub-cubes are considered to be door-to-door matches.}
\label{fig:matching_complexity}
\end{figure}

As alluded to in the introduction, door-to-door matching of complete trajectories from the origin to the destination is a structural obstacle to the transformation of carpooling to a mass transit service. To illustrate the difficulties of passenger-driver matching in space and in time for door-to-door trajectories, we can represent it with partition of a 3D cube divided into smaller sub-cubes, where the $x$-axis is the longitude, the $y$-axis the latitude and the $z$-axis the time, as shown in Figure~\ref{fig:matching_complexity}. On the left, there are 9 sub-cubes, where each sub-cube represents the origin/destination of a door-to-door trajectory.  The blue sub-cube in the lower left represents all the trajectories whose origins are, say, within a 5~km radius around a residential neighbourhood between 07:00 and 09:00 on Tuesday, and the green sub-cube the trajectories whose destination are within a 5~km radius of the workplace between 08:00 and 10:00 on Tuesday.  So for two trajectories to match spatio-temporally in a door-to-door sense, they must share the same sub-cube for the origin, and similarly for the destination: this condition is met by the 1 pair of green and blue sub-cubes among all possible 27 pairs of sub-cubes. On the right, the conditions for a door-to-door matching are stricter, say the origin is 1 km within the residential neighbourhood during 07:00 to 07:30, and the destination is 1 km within the workplace during 08:30 to 09:00. This represents 1 pair out of 125 pairs of sub-cubes. Thus stricter door-to-door matching leads to fewer drivers being available to share their trajectories with passengers.  

\begin{figure}[!ht]
\centering
\includegraphics[width=0.85\textwidth]{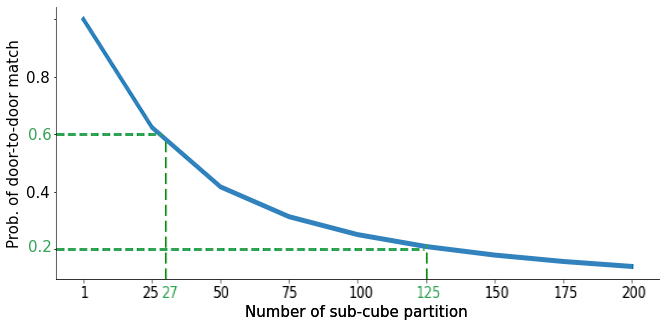}
\caption{Probability of door-to-door matches for uniformly distributed drivers and passengers, as a function of the number of sub-cube partition classes. Higher number of sub-cubes represent more stringent spatio-temporal matching conditions.}
\label{fig:matching_probs}
\end{figure}

To supplement the heuristic observations for door-to-door matching in Figure~\ref{fig:matching_complexity}, we demonstrate that the probability that two users (i.e. a driver and a passenger) share the same origin and destination at the same time decreases rapidly as the spatio-temporal matching conditions become more stringent. For the sake of simplicity, we suppose that the origin and destination for a driver and a passenger are both represented by independent random variables which are uniform over all sub-cubes in Figure~\ref{fig:matching_complexity}. Let $U_\mathrm{O}^{\mathrm{d}}$ and $U_\mathrm{D}^{\mathrm{d}}$ be the origin and destination of a driver, and likewise $U_\mathrm{O}^{\mathrm{p}}, U_\mathrm{D}^{\mathrm{p}}$ for a passenger. These quantities are all uniform random variables $\mathcal{U}(\{1, \dots, n\})$ where $n$ is the number of sub-cubes in Figure~\ref{fig:matching_complexity}. Then the probability of a door-to-door match between the driver and passenger is $p(n) = \mathbb{P} ( U_\mathrm{O}^\mathrm{p} = U_\mathrm{O}^\mathrm{d}, U_\mathrm{D}^\mathrm{p} = U_\mathrm{D}^\mathrm{d})$.  Since an exact formula for this probability is difficult to obtain, we approximate it by a Monte Carlo re-sampling method. That is, we generate $1000$ samples of $U_\mathrm{O}^\mathrm{p}, U_\mathrm{O}^\mathrm{d}, U_\mathrm{D}^\mathrm{p}, U_\mathrm{D}^\mathrm{d}$ and the probability of a door-to-door match is approximated as $$\hat{p} (n) = \frac{1}{1000} \sum_{i=1}^{1000} \pmb{1}\{U_{\mathrm{O}, i}^\mathrm{p} = U_{\mathrm{O},i}^\mathrm{d}, U_{\mathrm{D},i}^\mathrm{p} = U_{\mathrm{D},i}^\mathrm{d} \} $$ where $\pmb{1}\{\cdot\}$ is the indicator function. Figure~\ref{fig:matching_probs} is the graph of the number of sub-cubes $n$ versus the approximate probability of a door-to-door match $\hat{p}(n)$. If there is only 1 sub-cube (i.e. no spatio-temporal constraints) the probability of a match is 1. This probabilistic certainty decreases rapidly as the spatio-temporal constraints are added: for 27 sub-cubes, this probability is 0.6, and for 125 sub-cubes, it falls to 0.2. 

\begin{figure}[!ht]
\centering
\includegraphics[width=0.8\textwidth]{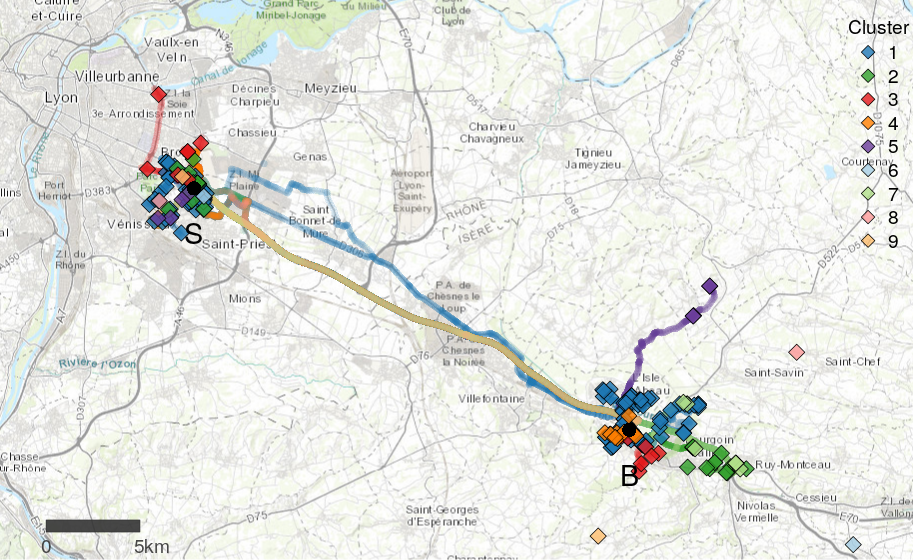} 
\caption{Spatio-temporal door-to-door matching fragments the number of mutualisable trajectories in an operational stochastic carpooling service. The clusters of GPS traces of door-to-door matches are colour coded, with the GPS points as the solid circles, and the origins/destinations as the solid diamonds. The meeting points are the solid black circles, denoted B = Bourgoin, S = St-Priest.}
\label{fig:lane-geotrack}
\end{figure}

The previous analysis was based on the synthetic uniformly distributed origins and destinations. For a more realistic example, we analyse some data generated by an operational carpooling service. Our example is the `Lane' carpooling service (\url{lanemove.com}) operated by Ecov, in conjunction with Instant System (\url{instant-system.com}), since May 2018 in the peri-urban regions around Lyon in south-eastern France. Our main data source is the GPS traces of drivers, which can be considered to be a form of crowd-sourced data collection \citep{Lee2011}. Passenger GPS traces are more difficult to obtain, and as we are not able to replicate exactly the synthetic example of passenger-driver matching above, so we use door-to-door matching of driver GPS traces to illustrate the diminishing probabilities. 
Since these GPS traces provide highly detailed spatio-temporal information, we are able to determine the number of strict door-to-door matches which also pass by two meeting points, as well as the number of matches when door-to-door matching is relaxed. For an illustrative example in Figure~\ref{fig:lane-geotrack}, we analyse 121 GPS traces of drivers who travelled from the Bourgoin meeting point (solid black circle labelled B) to the St-Priest meeting point (solid black circle labelled S) in the Lane carpooling service during the morning operating hours (06:30 to 09:00) for the work week 2019-11-25 to 2019-12-01. A hierarchical clustering with complete linkage was carried out on the spatial locations of these origins and destinations. The dissimilarity matrix used for this hierarchical clustering is composed of the Euclidean distance between the 4-vector comprising the (origin longitude, origin latitude, destination longitude, destination latitude) of each trajectory. This dissimilarity takes into account both the origin and the destination, but not the intermediate GPS points as these actual route taken is not critical for our purposes. We cut the dendrogram at $h=6000$ to yield 9 spatial clusters. These clusters are represented with the different colours. So GPS traces with the same colour can be considered as door-to-door matches with each other.  

The number of GPS traces per cluster is given in Table~\ref{tab:door-to-door-cluster}: as cluster 1 contains 75\% of the mutualisable traces, this leaves the other 25\% spread sparsely over the other 8 clusters, fragmenting the supply of the carpooling trajectories to passengers. 

\begin{table}[!ht]
\centering
\begin{tabular}{crrrrrrrrrrc}
\hline
Door-to-door cluster  & 1  & 2  & 3 & 4 & 5 & 6 & 7 & 8 & 9 & Total \\ 
Number of GPS traces   & 76 & 15 & 7 & 9 & 4 & 1 & 7 & 1 & 1 & 121 \\ \hline
\end{tabular}
\caption{Spatio-temporal door-to-door matching fragments the number of mutualisable trajectories in the Bourgoin \textgreater~St-Priest carpooling line, during its morning operating hours 06:30--09:00, from 2018-11-25 to 2018-12-01. The first line is the door-to-door cluster label and the second line is the number of traces in each cluster.}
\label{tab:door-to-door-cluster}
\end{table}

To quantify the augmentation of the carpooling potential by relaxing door-to-door matching, we compare the door-to-door cluster with the largest cardinality (76 traces) from Table~\ref{tab:door-to-door-cluster} to the number of the trajectories (121 traces) which coincide with this carpooling line regardless of their true origin and destination. These counts are an empirical equivalent of Figure~\ref{fig:matching_complexity}: the left corresponds to the 121 meeting point (i.e. relaxed door-to-door) matches, whereas the right the 76 door-to-door matches. Thus meeting point matching represents an increase of 45 traces or 59\% of the carpooling driver potential due to relaxing door-to-door matching.

Furthermore, \citet{stiglic2015} and \citet{xin2018plosone} provide more complex synthetic models to affirm that meeting points are essential to the feasibility of the mass carpooling services, and assert that it is almost impossible for a carpooling service to be based on door-to-door spatio-temporal matching. 

Whilst these examples demonstrate that incentivising drivers to converge to meeting points, rather than relying on door-to-door matching, increases the potential pool of mutualisable journeys, we have not yet demonstrated that this leads to reduced waiting times. This would be straightforward for the synthetic examples but this is not the case for empirically observed drivers and passengers journeys.  In the next section we introduce a general workflow which indeed allows us to confirm these reduced waiting times for empirical GPS traces.
 
\section{Data science-GIS workflow for the analysis of GPS traces}\label{sec:workflow}

\begin{figure}[!ht]
\centering
\includegraphics[width=0.95\textwidth]{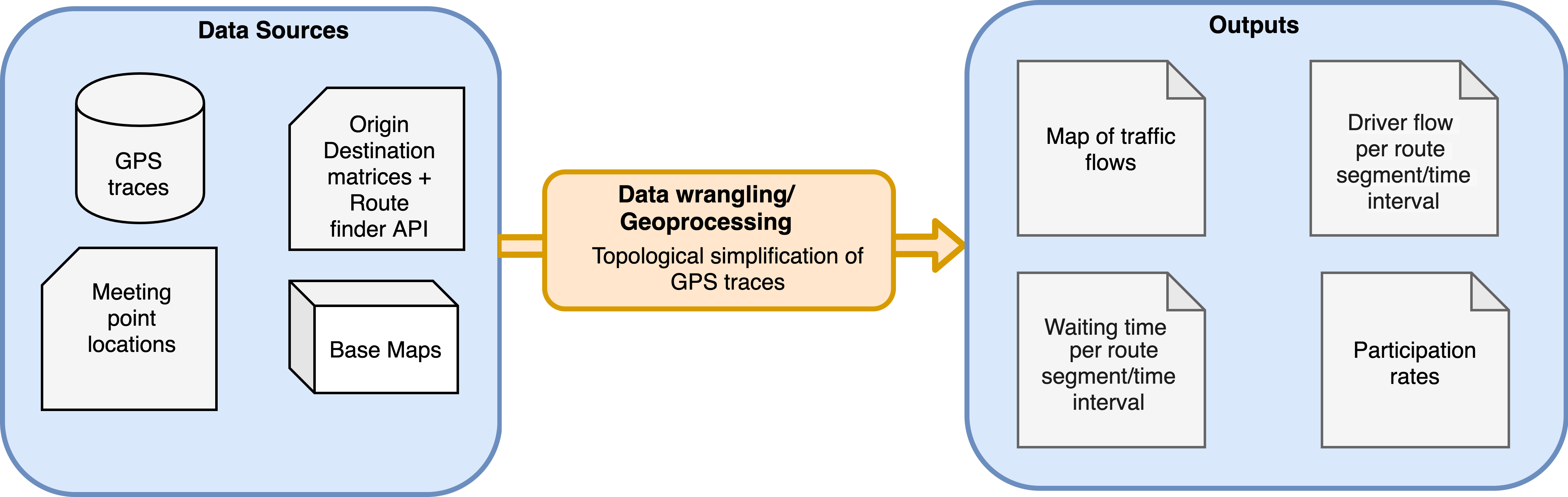}
\caption{Data science-GIS workflow for the analysis of driver GPS traces in stochastic carpooling service. (Left) Spatio-temporal input data sources. (Centre) Data wrangling and geoprocessing tasks. (Right) Generated outputs.}
\label{fig:flowchart}
\end{figure}

The GPS traces analysis workflow is illustrated in Figure~\ref{fig:flowchart}. The left rectangle of Figure~\ref{fig:flowchart} contains  the main data sources: the GPS traces, the meeting point locations, the origin-destination matrices, the route finder API and the base maps. The first two are supplied in-house by the carpooling service provider, the origin-destination matrices are usually supplied by a third party which has carried out a mobility survey (e.g. a national statistical agency \citet{mobpro}), the route finder API is provided by a GPS navigation operator (e.g. \citet{tomtom}), and the base maps are accessed from a cartography provider (e.g. \citet{osm}). There are specialised data wrangling techniques specific to spatial databases, known collectively as \textit{geoprocessing}, and these are carried out, in conjunction with traditional data wrangling, in the central rectangle. The critical geoprocessing task concerns the topological simplification of the GPS traces onto the carpooling network. Whilst GPS traces are a rich source of information of driver behaviour, they are voluminous and complex. Our approach is based on network analysis tools \citep{guidotti2017never} and complexity reduction/harmonisation algorithms \citep{Douglas2011}. This topological simplification is essential to be able to mutualise GPS traces which share common arrival times at the carpooling meeting points. Once these GPS traces are in a suitable format, we are able to produce the required outputs in the right rectangle, namely the predicted waiting times, the driver flow maps, the driver flow temporal profiles and the driver participation rates.  

\subsection{Data sources}

Our primary data source are the driver GPS traces. A GPS trace is represented by an $\ell$-sequence of triplets $\bX = \{ (X_i, Y_i, T_i) \}_{i=1}^\ell$ where $(X_i, Y_i)$ are the longitude, latitude coordinates of the GPS sensor at the $i^{th}$ timestamp $T_i$. We have $n$ GPS traces $\bX_1, \dots, \bX_n$ in the data collection period. The $m$ meeting point locations are represented by their GPS coordinates $\bM_1, \dots, \bM_m$. The origin-destination matrix is such that its $(j,k)^{th}$ entry is the number of journeys from $j^{th}$ origin to the $k^{th}$ destination.  In addition to the origin-destination matrix, we have the GPS coordinates of the origins and the destinations. Whilst it is common that they coincide, this is not required for our workflow. The base maps are graphics files of maps of the study area, which facilitate the fast and accurate map rendering at any desired scale.

\subsection{Data wrangling/geoprocessing}

From the $m$ meeting points $\bM_1, \dots, \bM_m$, a directed graph is constructed where the meeting points are the nodes, and an edge is drawn between the two nodes if carpooling between these two meeting points is guaranteed by the service provider. Thus a carpooling line is represented by an acyclic sub-graph with at least two nodes. 

The crucial data wrangling/geoprocessing process applied to the GPS traces is the topological simplification of GPS traces on a carpooling line. Around each of the $m$ meeting points, a buffer zone of 1 km radius is drawn to obtain $B(\bM_1), \dots, B(\bM_m)$. The intersection of the buffer zones and the GPS trace, $\bX \cap B(\bM_1), \dots, \bX \cap  B(\bM_m)$, is $m$ sub-sequences of the GPS points of $\bX$. For those meeting points with non-empty intersections, we consider that the driver is able to collect a passenger at these points without onerous detours. 

This spatial intersection only considers the spatial proximity of the driver to a passenger at a meeting point. For the carpooling to succeed, they also need to be in temporal proximity. Among the spatial intersections $\bX \cap B(\bM_1), \dots, \bX \cap  B(\bM_m)$, we examine the corresponding timestamps and retain only those in a suitably restrained time interval. If this reduced set of spatio-temporal intersections is non-empty then we proceed to the last data wrangling/geoprocessing step. 

We compute the closest GPS points in $\bX$ to the meeting points $\bM_j$, as defined by $\bX_{\bM_j} = \{ (X_k, Y_k, T_k) : k = \mathrm{argmin}_{1\leq i \leq \ell} \, \lVert ((X_i, Y_i) - \bM_j \lVert \}, j=1,\dots, m$. From this closest point $\bX_{\bM_j}$, we can extract the corresponding timestamp to be an estimate of the arrival time at the meeting point $\bM_j$. As an example, if the meeting points $\bM_1, \bM_2$ form the carpooling line $\{\bM_1 > \bM_2\}$, and if the GPS trace $\bX$ has well-defined estimated arrival times at $\bM_1$ and $\bM_2$, then we are able to reduce the complexity of the GPS trace. That is the $\ell$ points of $\bX$ can be reduced to the sequence of 4 points 
$$\tilde{\bX} (\bM_1, \bM_2) = \{ (X_1, Y_1, T_1) > \bX_{\bM_1} > \bX_{\bM_2} > (X_\ell, Y_\ell, T_\ell)\}$$ where $(X_1, Y_1, T_1)$ is the driver origin and $ (X_\ell, Y_\ell, T_\ell)$ is the driver destination. With this simplified trace $\tilde{\bX} (\bM_1, \bM_2)$, we are still able to determine if the driver can fulfil a passenger request at a given time on the carpooling line $\{\bM_1 > \bM_2\}$. The complex topology of $\bX$ is thus simplified by retaining a small number of key derived indicators \citep{Lee2011}.

We repeat these data wrangling/geoprocessing steps for all $n$ GPS traces. The result is a reduced set of $\tilde{n}$ GPS traces which correspond to the driver journeys which closely resemble the spatio-temporal characteristics of the likely passenger requests along the carpooling line.   

\subsection{Outputs} 

For the first output in the workflow in Figure~\ref{fig:flowchart}, if we visualise the GPS traces of the reduced set of $\tilde{n}$ meeting point matches with the base maps, then we obtain a map of the driver flow that matches to the passengers in the carpooling line, as in Figure~\ref{fig:lane-geotrack}.  
For the second output in the workflow, suppose that the initial time interval of interest is divided into $n_T$ sub-intervals $\tau_j, j =1, \dots, n_T$ since we wish to quantify the driver flow at a higher temporal resolution. Computing the driver flows $f(\tau_j), j = 1, \dots, n_T$, is straightforward as it only requires an enumeration of the  simplified GPS traces whose estimated arrival times fall within each sub-interval $\tau_j$. That is, the driver flow for the carpooling line $\{\bM_1 > \bM_2\}$ during the time interval $\tau_j$ is 
\begin{equation}
f (\tau_j,\bM_1, \bM_2) = \# \{ i : \tilde{\bX}_i (\bM_1, \bM_2) \in \tau_j, i= 1, \dots, \tilde{n}\}.
\label{eq:driver-flow}
\end{equation}

For the third output in the workflow, let $W(t)$ be the waiting time until the driver arrival for a carpool request made at time $t$. 
For stochastic carpooling, since a specific driver is not dispatched to the given passenger, the problem is equivalent to the arrival time of the first driver from the population of available drivers. Assuming a Poissonian driver arrival process, the waiting time and the driver flow are inversely proportional to each other, $W(t) \propto \mathrm{len}(\tau_j)/f_(\tau_j)$ where $t \in \tau_j$ and 
$\mathrm{len}(\tau_j)$ is the length of the time interval $\tau_j$. 
For simplicity, we set the constant of proportionality to 1 as this corresponds to the assumption that all geolocated drivers are willing to respond to a carpooling request. It is a reasonable assumption that the vast majority of geolocated drivers are willing to pick up a passenger, according to unpublished evidence supplied by Ecov. 
Thus for the carpooling line $\{\bM_1 > \bM_2\}$, the passenger waiting times for the time interval $\tau_j, j=1, \dots, n_T$, are
\begin{equation}
W(\tau_j,\bM_1, \bM_2) = \mathrm{len}(\tau_j)/f(\tau_j, \bM_1, \bM_2). 
\label{eq:wait-time}
\end{equation}

For the fourth output in the workflow, the driver participation rate is $P = n_1/n_0$ where $n_1$ is the total number of the drivers who are motivated to carpool in response to a passenger request, and $n_0$ is the total numbers of drivers who undertake journeys in the same geographical region as the carpooling service. Both $n_1$ and $n_0$ are difficult to define and to estimate precisely. We propose that $\tilde{n}$, calculated above as the number of drivers who share their geolocation, to be our proxy for $n_1$, as the vast majority of carpooling journeys are assured by drivers who are willing to share their geolocation. 

To enumerate all $n_0$ drivers in the same geographical region as the carpooling network is difficult since the GPS traces for all drivers are not available. Our proxy ($\tilde{n}_0)$ is derived from inferring likely trajectories from the reference origin-destination matrix. Usually this origin-destination matrix is provided at the county-level, but this is insufficiently detailed to decide if the drivers match with the meeting points on the carpooling lines. So we infer likely trajectories. These inferred likely trajectories are determined as the fastest route from the origins (county centroids) to the destinations (county centroids) by a route finder API. We employ a route finder API rather than an explicit model-based methodology, e.g. \citet{tang2016}, to infer these most likely routes. Model-based methods are the product of extensive theoretical and empirical work, and these tend to be difficult to access due to their proprietary nature. They also tend to be limited to dense urban regions, which are not the target regions for stochastic carpooling. Thus $\tilde{n}_0$ is the sum of the driver flow from all origin-destination pairs whose likely trajectories coincide with the carpooling lines. The driver participation rate for a carpooling line $\{\bM_1 > \bM_2\}$ is 
\begin{equation}
\tilde{P} = \tilde{n} / \tilde{n}_0
\label{eq:driver-participation}
\end{equation}
where $\tilde{n} = \sum_{j=1}^{n_T} f (\tau_j, \bM_1, \bM_2)$ from Equation~\eqref{eq:driver-flow}. 

Since there is no comparable door-to-door carpooling service operating concurrently with the meeting-point stochastic carpooling service, a direct comparison of empirical passenger waiting times is not possible. Instead, we propose an indirect comparison in three stages: (i) extract all driver GPS traces which connect two meeting points in a restrained time interval, as the meeting point matches, (ii) extract the largest hierarchical cluster of these GPS traces to serve as the door-to-door matches, and (iii) compute the driver flows using Equation~\eqref{eq:driver-flow} for both sets of matches, and then convert them using Equation~\eqref{eq:wait-time} to passenger waiting time predictions. 

In addition to the waiting times as an output, there are also the driver flow maps, the driver flow temporal profiles, and the driver participation rates. All these outputs are useful in understanding the transport mix of the local area as well as the market penetration of the carpooling into the transport mix. 

\section{Case study of an operational stochastic carpooling service} \label{sec:case}

\begin{figure}[!ht]
\centering
\includegraphics[width=0.75\textwidth]{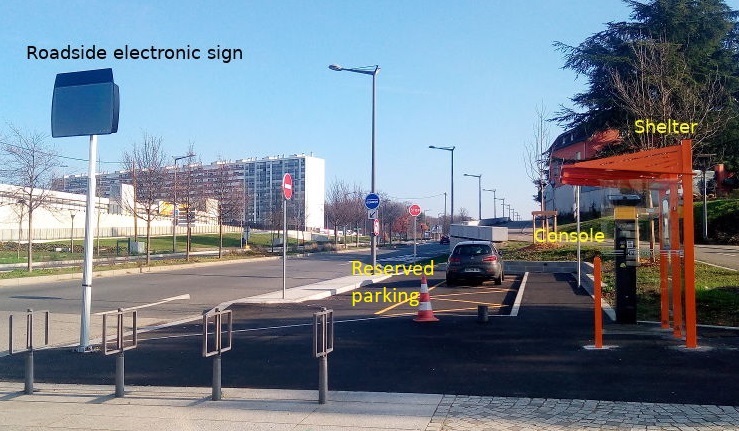}
\caption{Configuration of a physical meeting point for the `Lane' carpooling service. The orange structure is like a bus shelter. A passenger notifies potential drivers of their carpooling request using the console, which is then displayed on the roadside electronic sign. A driver can safely embark the passenger in the reserved parking place. Reproduced with permission from Ecov.}
\label{fig:lane-meeting-point}
\end{figure}

Our case study focuses on the Lane carpooling service introduced earlier. Before we progress further into the data analysis of the driver GPS traces, we describe the operational details of this stochastic carpooling service. The physical meeting points require an integrated infrastructure to facilitate this real-time stochastic matching, as illustrated in Figure~\ref{fig:lane-meeting-point}. The orange structure on the right functions like a bus shelter to provide protection from inclement weather whilst the passenger waits, and a prominent visual point of reference for drivers on the road. The passenger makes a carpooling request on the console (the green device with a horizontal yellow stripe). This request is displayed on the electronic sign on the roadside. In this configuration, the electronic sign is located close to the meeting point, but this can vary considerably according to the local geographical characteristics. A driver who wishes to embark the passenger in response to their request is able to do so safely in the reserved parking place.    

\subsection{Topological simplification of GPS traces on a carpooling line}

\begin{figure}[!ht]
\centering
\begin{tabular}{@{}c@{}c@{}}
\includegraphics[width=0.5\textwidth]{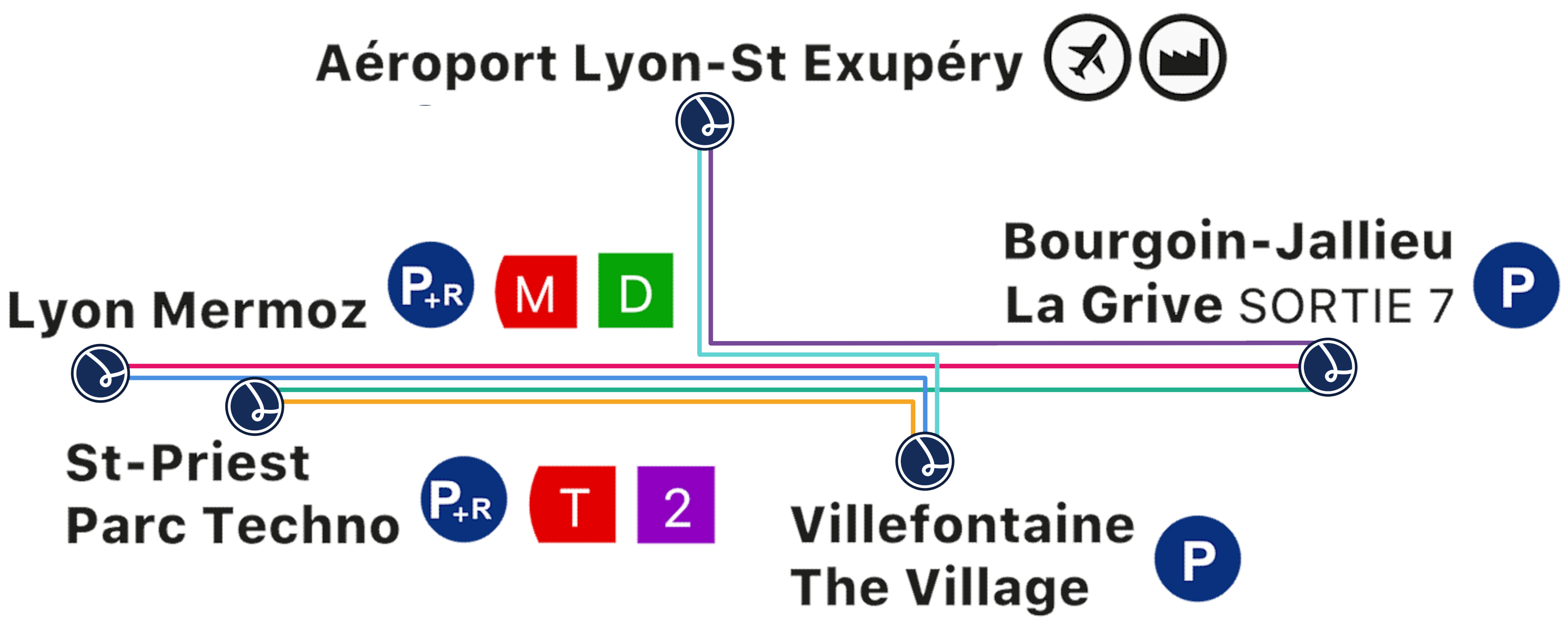} &
\includegraphics[width=0.49\textwidth]{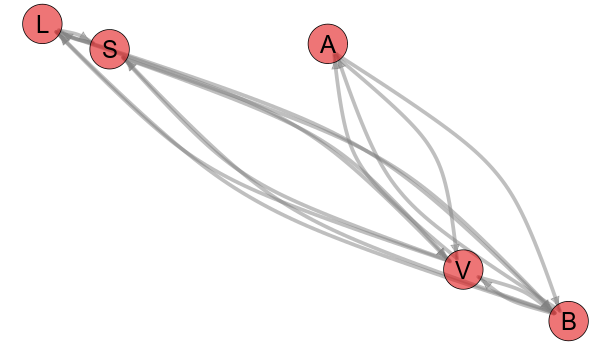}  
\end{tabular}
\caption{(Left) Schematic diagram of the Lane carpooling network, which resembles the geographically restrained trajectories of a public transport service. Reproduced with permission from Ecov. (Right) Carpooling network represented as a directed graph. Nodes are the meeting points, edges connect meeting points whenever a carpooling service between them is assured.}
\label{fig:lane-graph}
\end{figure}

The schematic diagram of the carpooling lines in the Lane network is shown on the left of Figure~\ref{fig:lane-graph}. The visual similarities of the schematic of this carpooling service with those associated with bus or train services is deliberately designed to induce the perception of carpooling as a form of public transport. There are 5 physical meeting points (Lyon Mermoz, St-Priest Parc Techno, Aéroport Lyon-St Exupéry, Villefontaine The Village, and Bourgoin La Grive Sortie 7), denoted by the circles with the stylised $\mathcal{L}$, which function analogously to bus stops. According to mobility studies in this territory, the coloured lines connect the meeting points that have a sufficient driver flow between them to maintain a carpooling service with stochastic matching. These connected meeting points form a carpooling line, analogous to a bus line, where carpooling is only available between these meeting points.

This carpooling network is represented as a directed graph, as shown on the right of  Figure~\ref{fig:lane-graph}, where each node is a meeting point and the edge connects two nodes if they form segment of a carpooling line. For brevity, the node labels are abbreviated to the first letter, i.e. L = Lyon Mermoz, S = St-Priest Parc Techno, A = Aéroport Lyon-St Exupéry, V = Villefontaine The Village, and B = Bourgoin La Grive Sortie 7. We focus on the most frequented carpooling line, that is, the Bourgoin \textgreater~St-Priest line (green line in Figure~\ref{fig:lane-graph}). The topology of the road network ensures that most of the journeys from Bourgoin to St-Priest pass also by the Villefontaine meeting point, that is, the B \textgreater~S carpooling line includes both sub-graphs B \textgreater~S and B \textgreater~V \textgreater~S. The period of data collection is 2019-07-25 (service launch date) to 2020-02-17 (last date for which consistent driver GPS traces are available), during the most frequented time period (the morning operating hours 06:30-09:00).

\begin{figure}[!ht]
\centering
\includegraphics[width=0.8\textwidth]{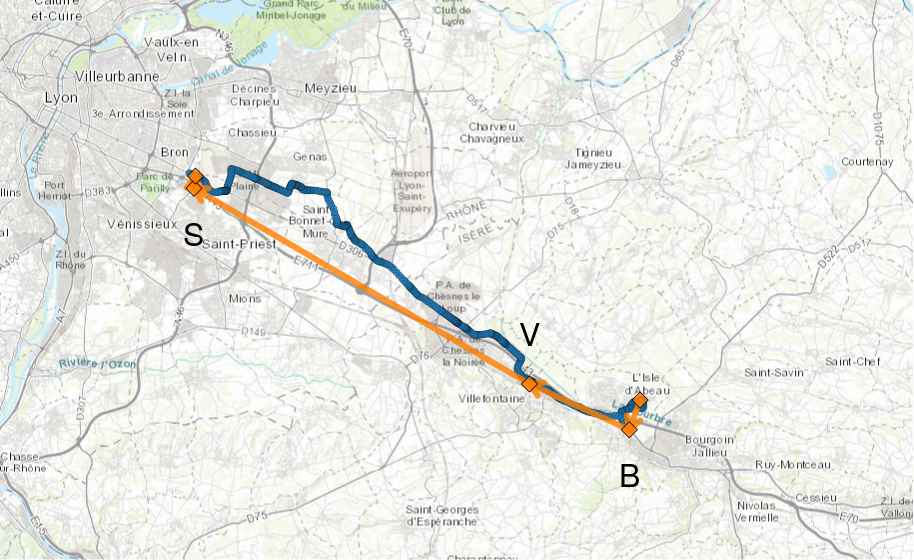}
\caption{Topological simplification of a GPS trace in the Bourgoin \textgreater~St-Priest carpooling line, during its morning operating hours 06:30--09:00, on 2019-11-28. The complete GPS trace are the 530 blue circles; the sequence of five nodes, as its simplified topology, are the orange arrows, and the orange diamonds are the origin, carpooling meeting points, destination nodes. The meeting points are S = St-Priest, V = Villefontaine, and B = Bourgoin.} 
\label{fig:lane-route}
\end{figure}

A complete GPS trace $\bX$ is displayed as the sequence of 530 blue circles in Figure~\ref{fig:lane-route}. This GPS trace passes within 1 km of the B, V and S meeting point nodes, so its simplified topology consists of the 5-node sequence \{origin \textgreater~B \textgreater~V \textgreater~S \textgreater~destination\}, shown as the orange arrows.  

This simplified GPS trace represents a data compression rate of over 99\% yet it retains the important information to decide the matching potential of this GPS driver trace with a passenger request on the Bourgoin \textgreater~St-Priest carpooling line. In Table~\ref{tab:lane-geotrack-compression} is the average data compression for the $\tilde{n}=121$ GPS traces on the Bourgoin \textgreater~St-Priest line. The first column is the average number of GPS points in the complete driver traces $\#\bX$, the second is the average number of GPS points of the simplified topologies $\#\tilde{\bX}$, and the last column is the average data compression rate $(1- \#\tilde{\bX}/ \#\bX)$.

\begin{table}[!ht]
\centering
\begin{tabular}{cccc}
\hline
Line & \makecell{\#\,points in complete\\GPS traces} & \makecell{\#\,points in simplified\\GPS traces} & \%\,compression\\
B \textgreater~S & 313 & 5 & 98.3  \\
\hline
\end{tabular}
\caption{Data compression rate on the Bourgoin \textgreater~St-Priest carpooling line, during the morning operating hours 06:30--09:00, for all driver GPS traces from 2019-07-25 to 2020-02-17. The first column is the average number of GPS points in the complete driver traces, the second is the average number of points of the simplified GPS traces, and the third column is the average data compression rate.}
\label{tab:lane-geotrack-compression}
\end{table}

\subsection{Driver flow estimation} 

\begin{figure}[!ht]
\centering
\includegraphics[width=0.8\textwidth]{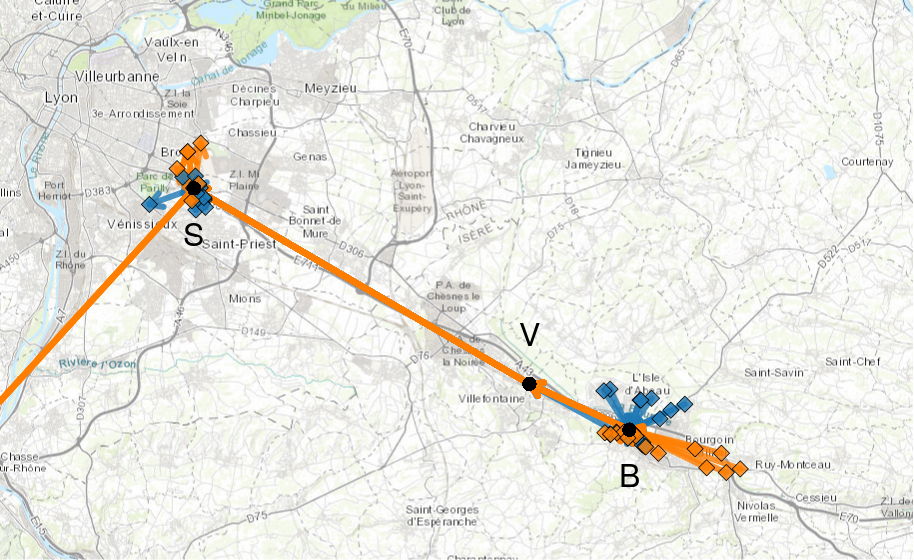} 
\caption{Matching on meeting points increases the number of driver spatio-temporal matches in comparison to door-to-door matching, during a single 30 minute period (08:00-08:30), on 2019-11-28. The orange arrows are the 14 GPS traces  which are both meeting point and door-to-door, and the blue arrows are the 17 GPS traces which are  meeting point matches but not door-to-door match matches. The diamonds are the origin/destination points. The solid black circles are the meeting points: S = St-Priest, V = Villefontaine, and B = Bourgoin.} 
\label{fig:lane-route2}
\end{figure}

Of the $\tilde{n}=121$ GPS traces that follow the  Bourgoin \textgreater~St-Priest carpooling line, $31$ GPS traces have an arrival time at Bourgoin within 08:00 to 08:30, i.e., a close spatio-temporal match for a passenger request for a departure at the Bourgoin meeting point between 08:00 and 08:30 am, with a destination at the St-Priest meeting point. Of these 31 GPS traces, 17 of them are door-to-door matches (as defined as belonging to the largest door-to-door cluster of GPS traces in Table~\ref{tab:door-to-door-cluster}). These 17 traces are both door-to-door and meeting point matches and their simplified traces are displayed in Figure~\ref{fig:lane-route2} as the orange diamonds/arrows. The simplified traces of the remaining 14 meeting point but not door-to-door matches are the blue diamonds/arrows. The latter represent an 82\% increase in the number of drivers (from 17 to 31) who can potentially respond to a passenger carpooling request on the Bourgoin \textgreater~St-Priest line. 

Table~\ref{tab:lane-geotrack-flow-increase} summarises the weekly evolution of the impact of meeting point matching over door-to-door matching. The first set of three columns focus on the entire morning operating hours (06:30--09:00) whereas the second set on the single 30 minute period (08:00--08:30) as this latter restricted period is a more realistic time frame that potential passengers are willing to wait for a driver to arrive. The first column contains the weekly aggregate number of door-to-door matches, the second the number of meeting point matches, and the third is the percentage increase due to meeting point matches, i.e. $(\#\,\textrm{meeting point matches} - \#\,\textrm{door-to-door matches})/\#\,\textrm{door-to-door matches}$. The number of door-to-door matches are enumerated from a similar hierarchical clustering to that in  Table~\ref{tab:door-to-door-cluster}, and the number of meeting point matches are computed from Equation~\eqref{eq:driver-flow}. 
This table demonstrates that the increase in the driver flow due to meeting point matching is maintained over the entire data collection period. 

\begin{table}[!ht]
\centering
\begin{tabular}{crrrrrr}
\hline
& \multicolumn{3}{c}{Driver flow (06:30--09:00)} & \multicolumn{3}{c}{Driver flow (08:00--08:30)} \\
Week &  \makecell{\#\,Door-to-\\door} & \makecell{\#\,Meeting\\point} & \%\,increase  & \makecell{\#\,Door-to-\\door} & \makecell{\#\,Meeting\\point} & \%\,increase \\ 
 2019W36 & 54 & 100 & 85 & 18 & 24 & 33 \\
 2019W37 & 67 & 99  & 48 & 20 & 21 & 5 \\
 2019W38 & 81 & 122 & 51 & 20 & 27 & 35 \\
 2019W39 & 72 & 119 & 65 & 20 & 28 & 40 \\
 2019W40 & 43 & 94  & 119 & 9 & 19 & 111 \\
 2019W41 & 48 & 106 & 120 & 12 & 29 & 141 \\
 2019W42 & 50 & 103 & 106 & 18 & 33 & 83 \\
 2019W43 & 30 & 85  & 183 & 11 & 27 & 145 \\
 2019W44 & 43 & 63  & 47 & 9 & 14 & 56 \\
 2019W45 & 48 & 102 & 113 & 14 & 28 & 100 \\
 2019W46 & 41 & 71  & 73 & 12 & 22 & 83 \\
 2019W47 & 60 & 110 & 83 & 15 & 30  & 100 \\
 2019W48 & 76 & 121 &  59 & 17 & 31 & 82 \\
 2019W49 & 58 & 94 & 62 & 19 & 32 & 68 \\
 2019W50 & 47 & 99 & 111 & 5 & 22 & 340 \\
 2019W51 & 82 & 103 & 26 & 21 & 27 & 29 \\
 2020W01 & 12 & 23 & 92 & 4 & 6 & 50 \\
 2020W02 & 53 & 96 & 81 & 14 & 23 & 64 \\
 2020W03 & 63 & 100 & 59 & 14 & 27 & 93 \\
 2020W04 & 74 & 105 & 42 & 16 & 23 & 44 \\
 2020W05 & 55 & 104 & 89 & 16 & 23 & 44 \\
 2020W06 & 44 & 110 & 150 & 14 & 27 & 93 \\
 2020W07 & 57 & 96 & 68 & 13 & 23 & 77 \\
\hline
\end{tabular}
\caption{Weekly aggregate driver flow increase of meeting point matching compared to door-to-door matching in the Bourgoin \textgreater~St-Priest carpooling line, during the morning operating hours 06:30--09:00, and 08:00--08:30, from 2019-09-02 to 2020-02-17. The first columns contains the number of door-to-door matches, the second the number of meeting point matches, and the third the percentage increase due to the meeting point matches.}
\label{tab:lane-geotrack-flow-increase}
\end{table}

The simplified GPS traces also allow us to compute the driver flows for narrower time intervals than the 2.5 hour and 0.5 hour intervals in Table~\ref{tab:lane-geotrack-flow-increase}. Following \citet{smith1997traffic} and \citet{mcshane1990traffic} that 15 minutes intervals are a suitable choice because the variation in driver flows for shorter intervals is less stable,   Table~\ref{tab:lane-geotrack-flow2} displays the average driver flow for 15 minute intervals during 06:30 to 09:00. For robustness, we aggregate these driver flows over all weeks in the data collection period in applying Equation~\eqref{eq:driver-flow} since we have increased the intra-day temporal resolution. 

\begin{table}[!ht]
\centering
\begin{tabular}{@{}crrrrrrrrrr@{}}
\hline
& \multicolumn{10}{c}{Driver flow}  \\
Line    & \makecell{06:30\\--06:45} & \makecell{06:45\\--07:00} & \makecell{07:00\\--07:15} & \makecell{07:15\\--07:30} & \makecell{07:30\\--07:45} & \makecell{07:45\\--08:00} & \makecell{08:00\\--08:45} & \makecell{08:15\\--08:30} & \makecell{08:30\\--08:45} & \makecell{08:45\\--09:00}  \\
B \textgreater~S & 1 & 1.5 & 2.5 & 1.5 & 3 & 1.5 & 2 & 2 & 2 & 1\\ 
\hline
\end{tabular}
\caption{Daily average driver flow on the Bourgoin \textgreater~St-Priest carpooling line, per 15 minute intervals, during the morning operating hours 06:30--09:00 for a typical day from 2019-09-02 to 2020-02-17.}
\label{tab:lane-geotrack-flow2}
\end{table}

\subsection{Waiting time prediction}

It is straightforward to convert these average driver flows in Table~\ref{tab:lane-geotrack-flow2} into predicted waiting times using Equation~\eqref{eq:wait-time}. Suppose that a passenger makes a carpool request at 08:10 at the Bourgoin meeting point to travel to St-Priest. The expected waiting time is the length of the interval divided by the average driver flow in the interval 08:00--08:15, i.e. 7.5 minutes.  Given that we have already established a highly detailed spatio-temporal profile of the average driver flow for a carpooling line in Table~\ref{tab:lane-geotrack-flow2}, then the predicted waiting times at the same temporal resolution are shown in Table~\ref{tab:lane-geotrack-wait}.  

\begin{table}[!ht]
\centering
\begin{tabular}{@{}crrrrrrrrrr@{}}
\hline
& \multicolumn{10}{c}{Predicted waiting time (min)}  \\
Line    & \makecell{06:30\\--06:45} & \makecell{06:45\\--07:00} & \makecell{07:00\\--07:15} & \makecell{07:15\\--07:30} & \makecell{07:30\\--07:45} & \makecell{07:45\\--08:00} & \makecell{08:00\\--08:45} & \makecell{08:15\\--08:30} & \makecell{08:30\\--08:45} & \makecell{08:45\\--09:00}  \\
B \textgreater~S & 15.0 & 10.0 & 6.0 & 10 & 6.0 & 5.0 & 7.5 & 7.5 & 7.5 & 15 \\ 
\hline
\end{tabular}
\caption{Waiting time predictions for a carpool request on the Bourgoin \textgreater~St-Priest carpooling line, per 15 minute intervals, during the morning operating hours 06:30--09:00 for a typical day from 2019-09-02 to 2020-02-17.}
\label{tab:lane-geotrack-wait}
\end{table}

\begin{figure}[!ht]
\centering
\includegraphics[width=0.85\textwidth]{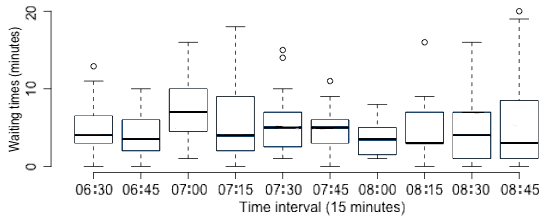} 
\caption{Box plots of observed waiting times on the Bourgoin \textgreater~St-Priest carpooling line, per 15 minute intervals during the morning operating hours 06:30--09:00, from 2019-09-02 to 2020-02-17.} 
\label{fig:waiting_times}
\end{figure}

For the Bourgoin \textgreater~St-Priest carpooling line from 2019-07-25 to 2020-02-17, we observed roughly 1500 carpooling requests with a reliably recorded waiting time. Each box plot in Figure~\ref{fig:waiting_times} displays the observed waiting times each 15 minute interval during the morning opening hours with at least one observed waiting time.

\begin{figure}[!ht]
\centering
\includegraphics[width=0.85\textwidth]{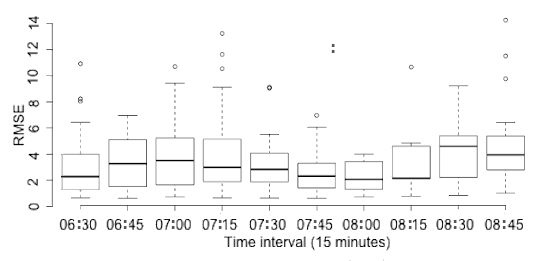} 
\caption{RMSE between the observed and predicted waiting times on the Bourgoin \textgreater~St-Priest carpooling line, per 15 minute intervals during the morning operating hours 06:30--09:00, from 2019-09-02 to 2020-02-17.} 
\label{fig:rmse_interval}
\end{figure}

The accuracy of these predicted waiting times with respect to these observed ones is illustrated in Figure~\ref{fig:rmse_interval}. Each box plot displays the RMSE (Root Mean Squared Error) between the observed and the predicted waiting times (from Table~\ref{tab:lane-geotrack-wait}) for each 15 minute interval. For all 15 minute intervals, the median RMSE is around 2 to 4 minutes which implies that the predicted waiting times fairly closely track the observed waiting times. This gives us confidence in our method for predicting waiting times in a stochastic carpooling service.

Since there is no comparable operational door-to-door carpooling service to Bourgoin \textgreater~St-Priest stochastic carpooling line, then it is not possible to compare empirical waiting times from each type of carpooling. Our proxy is to compare the predicted waiting times for the door-to-door matches and the meeting point matches. Since our method for waiting time prediction is fairly accurate for meeting point matches according to Figure~\ref{fig:rmse_interval}, we reason that it will also yield accurate waiting times for door-to-door matches. In Table~\ref{tab:lane-geotrack-waiting-time-decrease} are the predicted waiting times based on the weekly aggregate driver flows from Table~\ref{tab:lane-geotrack-flow-increase} via Equation~\eqref{eq:wait-time} for both door-to-door and meeting point matches. For all weeks, we observe a decrease in the predicted passenger waiting times. From anecdotal evidence from Ecov, 15 minutes corresponds roughly to the maximum time that passengers are willing to wait for a driver to arrive since a pre-arranged meeting time has not made. This 15 minute threshold is exceeded by the door-to-door waiting times for most weeks, whereas the meeting point matched waiting times are lower than 15 minutes for most weeks.

\begin{table}[!ht]
\centering
\begin{tabular}{@{}crrrrrr@{}}
\hline
& \multicolumn{3}{c}{Predicted waiting time (06:30--09:00)} & \multicolumn{3}{c}{Predicted waiting time (08:00--08:30)} \\
Week &  \makecell{Door-to-\\door (min) } & \makecell{Meeting\\point (min)} & \%\,decrease  & \makecell{Door-to-\\door (min)} & \makecell{Meeting\\point (min)} & \%\,decrease \\ 
 2019W36 & 13.9 & 7.5 & --46 & 8.3 & 6.2 & --25 \\
 2019W37 & 11.2 & 7.6 & --32 & 7.5 & 7.1 & --5 \\
 2019W38 &  9.3 & 6.1 & --34 & 7.5 & 5.6 & --26 \\
 2019W39 & 10.4 & 6.3 & --39 & 7.5 & 5.4 & --29 \\
 2019W40 & 17.4 & 8.0 & --54 &16.7 & 7.9 & --53 \\
 2019W41 & 15.6 & 7.1 & --55 &12.5 & 5.2 & --59 \\
 2019W42 & 15.0 & 7.3 & --51 & 8.3 & 4.5 & --45 \\
 2019W43 & 25.0 & 8.8 & --65 &13.6 & 5.6 & --59 \\
 2019W44 & 17.4 &11.9 & --32 &16.7 &10.7 & --36 \\
 2019W45 & 15.6 & 7.4 & --53 &10.7 & 5.4 & --50 \\
 2019W46 & 18.3 &10.6 & --42 &12.5 & 6.8 & --45 \\
 2019W47 & 12.5 & 6.8 & --45 &10.0 & 5.0 & --50 \\
 2019W48 &  9.9 & 6.2 & --37 & 8.8 & 4.8 & --45 \\
 2019W49 & 12.9 & 8.0 & --38 & 7.9 & 4.7 & --41 \\
 2019W50 & 16.0 & 7.6 & --53 &30.0 & 6.8 & --77 \\
 2019W51 &  9.1 & 7.3 & --20 & 7.1 & 5.6 & --22 \\
 2020W01 & 14.2 & 7.8 & --45 &10.7 & 6.5 & --39 \\
 2020W02 & 11.9 & 7.5 & --37 &10.7 & 5.6 & --48 \\
 2020W03 & 10.1 & 7.1 & --30 & 9.4 & 6.5 & --30 \\
 2020W04 & 13.6 & 7.2 & --47 & 9.4 & 6.5 & --30 \\
 2020W05 & 17.0 & 6.8 & --60 &10.7 & 5.6 & --48 \\
 2020W06 & 13.2 & 7.8 & --41 &11.5 & 6.5 & --43 \\
 2020W07 & 53.6 &41.7 & --22 &30.0 &30.0 & 0 \\
\hline
\end{tabular}
\caption{Weekly predicted passenger waiting times for door-to-door and meeting point matching in the Bourgoin \textgreater~St-Priest carpooling line, during the morning operating hours 06:30--09:00, and 08:00--08:30, from 2019-09-02 to 2020-02-17. The first columns contains the predicted waiting times for door-to-door matches, the second for  meeting point matches, and the third the percentage decrease due to the meeting point matches.}
\label{tab:lane-geotrack-waiting-time-decrease}
\end{table}

\subsection{Driver participation rate estimation}

\begin{figure}[!ht]
\centering
\includegraphics[width=0.8\textwidth]{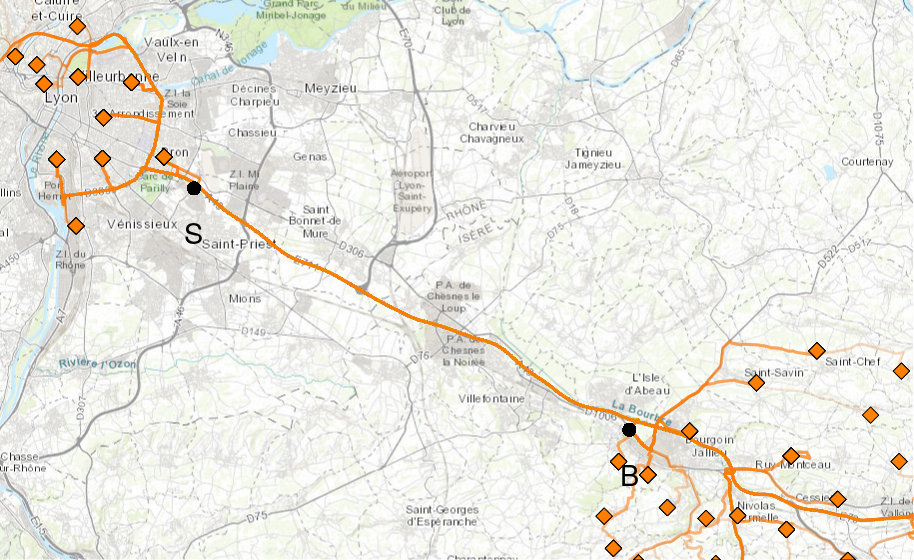} 
\caption{Likely driver itineraries from the TomTom route finder API in the same geographical region as the Bourgoin \textgreater~St-Priest carpooling line. The origins and destinations (county centroids) are the orange diamonds. The solid black circles are the meeting points: S = St-Priest, B = Bourgoin.} 
\label{fig:mobpro}
\end{figure}

A key question for the service provider is what driver participation rate leads to passenger waiting times around 5 to 10 minutes, as observed in Figure~\ref{fig:waiting_times}?  To respond to this question, we first need to enumerate the population of all drivers on a carpooling line. The county-level origin-destination matrix of home-work trajectories from the French official statistical agency \citep{mobpro} is insufficiently detailed to decide if the drivers with these origins-destinations travel on the carpooling lines.  So we infer likely trajectories, as determined as the fastest route by the TomTom route finder API \citep{tomtom} starting on Tuesday 8am  from the origins (county centroids) to the destinations (county centroids). A spatial intersection, similar to that carried out for the driver GPS traces, is computed to determine which trajectories pass within 1 km of the carpooling meeting points. These trajectories are shown in Figure~\ref{fig:mobpro}. Note that there is no temporal information attached to this origin-destination matrix, but since they are home-work trajectories, we suppose they are effected in the morning peak hours which matches the time interval of the driver GPS traces. 

If we then aggregate the corresponding driver flows in the origin-destination matrix, then we obtain  $\tilde{n}_0 = 3821$ drivers whose likely trajectories match the Bourgoin \textgreater~St-Priest carpooling line. From Table~\ref{tab:lane-geotrack-flow2}, there is a daily average of $\tilde{n}=20$ driver GPS traces between 06:30 and 09:30. This yields an estimated driver participation rate of $\tilde{P}=\tilde{n}/\tilde{n}_0=0.52\%$ from Equation~\eqref{eq:driver-participation}. Even with this low driver participation rate, average passenger waiting times of 5--10 minutes are observed in Figure~\ref{fig:waiting_times}.

\begin{figure}[!ht]
\centering
\includegraphics[width=0.75\textwidth]{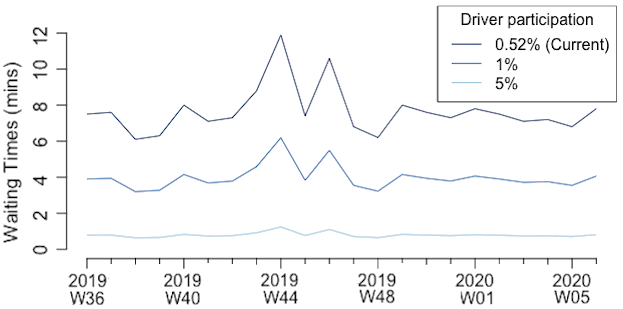} 
\caption{Evolution of the predicted passenger waiting time as a function of the driver participation rate in the Bourgoin \textgreater~St-Priest carpooling line, during the morning operating hours 06:30--09:00, from 2019-07-25 to 2020-02-17.} 
\label{fig:lane-geotrack-participation}
\end{figure}

If we were able to increase this low driver participation rate even modestly (to 1\% or 5\%), then the predicted passenger waiting times would fall substantially, as illustrated in Figure~\ref{fig:lane-geotrack-participation}. In this case, these waiting times would be lower than those of bus lines and approach those of high frequency metro/subway train lines. The methods for increasing driver participation, as they lie largely outside of data science, are out of scope of this paper but are of intense interest to the service provider \citep{zhu2017more,zhu2018,zhu2020}.  

\section{Conclusions and future work}

Stochastic real-time carpooling services differ from competing services which offer deterministic door-to-door matching for complete trajectories. Whilst the latter offer a high level of personal convenience in highly urbanised regions, door-to-door matching structurally inhibits mass adoption of carpooling, especially in peri-urban regions. Relaxing the strict door-to-door matching allows, and subsequently implementing stochastic meeting point matching, allows for the mutualisation of high throughput road segments, and thus removes this obstacle. We introduced a novel data science-GIS workflow for a stochastic carpooling service. The crucial mathematical abstraction in this workflow is to reduce the complexity of driver GPS traces to a graph-based topology of the carpooling network. We illustrated this workflow on an operational stochastic carpooling service in a peri-urban region in south-eastern France. We provided quantitative justifications that the physical meeting points, by facilitating a critical mass of drivers and passengers drawn from a much larger geographical area, leads to passenger waiting times which are lower than those achieved by door-to-door matching. Our workflow is novel combination of two closely related, but historically separate, disciplines of data science and GIS into a single workflow. In addition to predicting the passenger waiting times, it is able to deliver outputs for the driver flow maps, driver flow spatio-temporal profiles, and driver participation rates. This workflow forms a solid prototype for other workflows to accompany the expansion of stochastic carpooling services to address the mobility requirements in neglected peri-urban regions in the future.

\section{Acknowledgements}
The authors thank Ecov for providing the data sets of driver GPS traces and passenger waiting times. The authors also thank Bertrand Michel from the Central Engineering School of Nantes and Gérard Biau from Sorbonne University for their feedback.

\bibliographystyle{plainnat}
\bibliography{references}

\end{document}